\documentclass[12pt]{article}
\usepackage{graphicx}

\newcommand\pubnumber{WSU--HEP--XXYY}
\newcommand\pubdate{\today}

\def\uhhi{Department of Physics and Astronomy\\
University of Hawaii, Honolulu, HI 96822, USA}
\def\uhsupport{\footnote{Work supported by the DOE grant: Research in High Energy Physics, DESC-0010504.}}

\def\Title#1{\begin{center} {\Large #1 } \end{center}}
\def\Author#1{\begin{center}{ \sc #1} \end{center}}
\def\Address#1{\begin{center}{ \it #1} \end{center}}

\newcommand\pubblock{\rightline{\begin{tabular}{l} \pubnumber\\
         \pubdate  \end{tabular}}}
\newenvironment{Abstract}{\begin{quotation}  }{\end{quotation}}
\newenvironment{Presented}{\begin{quotation} \begin{center} 
             PRESENTED AT\end{center}\bigskip 
      \begin{center}\begin{large}}{\end{large}\end{center} \end{quotation}}
\def\Acknowledgements{\bigskip  \bigskip \begin{center} \begin{large}
             \bf ACKNOWLEDGEMENTS \end{large}\end{center}}

\def\beq{\begin{equation}}
\def\eeq#1{\label{#1}\end{equation}}
\def\eeqn{\end{equation}}

\def\beqa{\begin{eqnarray}}
\def\eeqa#1{\label{#1}\end{eqnarray}}
\def\eeqan{\end{eqnarray}}

\let\bar=\overbar

\def\Dslash{\not{\hbox{\kern-4pt $D$}}}
\def\dslash{\not{\hbox{\kern-2pt $\del$}}}

\def\msb{{\bar{\ssstyle M \kern -1pt S}}}

\newcommand{\overl}[1]{\mkern 1.5mu\overline{\mkern-3.5mu#1\mkern-1.5mu}\mkern 1.5mu}

\begin{document}
\begin{titlepage}
\pubblock

\vfill
\Title{Charmonia and exotics from BESIII}

\vfill
\Author{Mihajlo Kornicer\uhsupport}
\Address{\uhhi}
\vfill
\begin{Abstract}Several states that couple to one light-meson and a heavy-charmonium are recently reported,
including the $Z_c(3900)$ and $Z_c(4020)$. These states 
cannot be interpreted within  the constituent quark-model of mesons, and are good candidates for a 
new form of quark matter, four-quark matter. A review of recent results on new charmonium-like states
studied at BESIII is presented.
\end{Abstract}
\vfill
\begin{Presented}
The 7th International Workshop on Charm Physics (CHARM 2015)\\
Detroit, MI, 18-22 May, 2015
\end{Presented}
\vfill
\end{titlepage}
\def\thefootnote{\fnsymbol{footnote}}
\setcounter{footnote}{0}
%

\section{Introduction}

The quark model~\cite{Fritzsch:1973pi}, which treats mesons as combinations of one quark and one anti-quark, 
is very successful in describing meson properties, particularly in the charmonium ($c \bar{c}$) 
region below the open-charm threshold.~\cite{Barnes:2005pb}.
However, within the QCD and strongly coupled gauge theories~\cite{Brambilla:2014jmp}, 
we can expect structures that are more than just  $q \bar{q}$ combinations. 
Recent discoveries of heavy-flavor mesons~\cite{Brambilla:2010cs}  that do not fit 
into this simple quark-anti-quark picture, labeled $XYZ$, has inspired 
particle physics community to intensify studies to understand the nature and properties of these states.


\section{Z-states studied by BESIII}

The BESIII experiment was first to report the $Z_c(3900)$ state~\cite{Ablikim:2013mio} 
in the $\pi J/\psi$ final state, 
and soon the $Z_c(4020)$~\cite{Ablikim:2013wzq} was observed in the $\pi h_c$ decays. 
Both of these states were discovered as charged particles, coupled to a pion and charmonium,  
making them four-quark ($u \bar{d} c \bar{c}$) meson candidates. 
However, both states have been also reported in respective neutral modes, 
thus complementing the isospin-triplet representation of isospin one, $I=1$, resonances. As these two states 
are close to $D \overl{D}^{\star}$ and $D^{\star} \overl{D}^{\star}$ thresholds, respectively, 
new structures coupling to two charmed mesons  were soon discovered.
The BESIII experiment is collecting more data in the $XYZ$ region, to map the properties 
and decay modes of newly discovered states.

\subsection{ $Z_c \to \pi J/\psi$}

\begin{figure}[htb]
\centering
\includegraphics[height=2.5in]{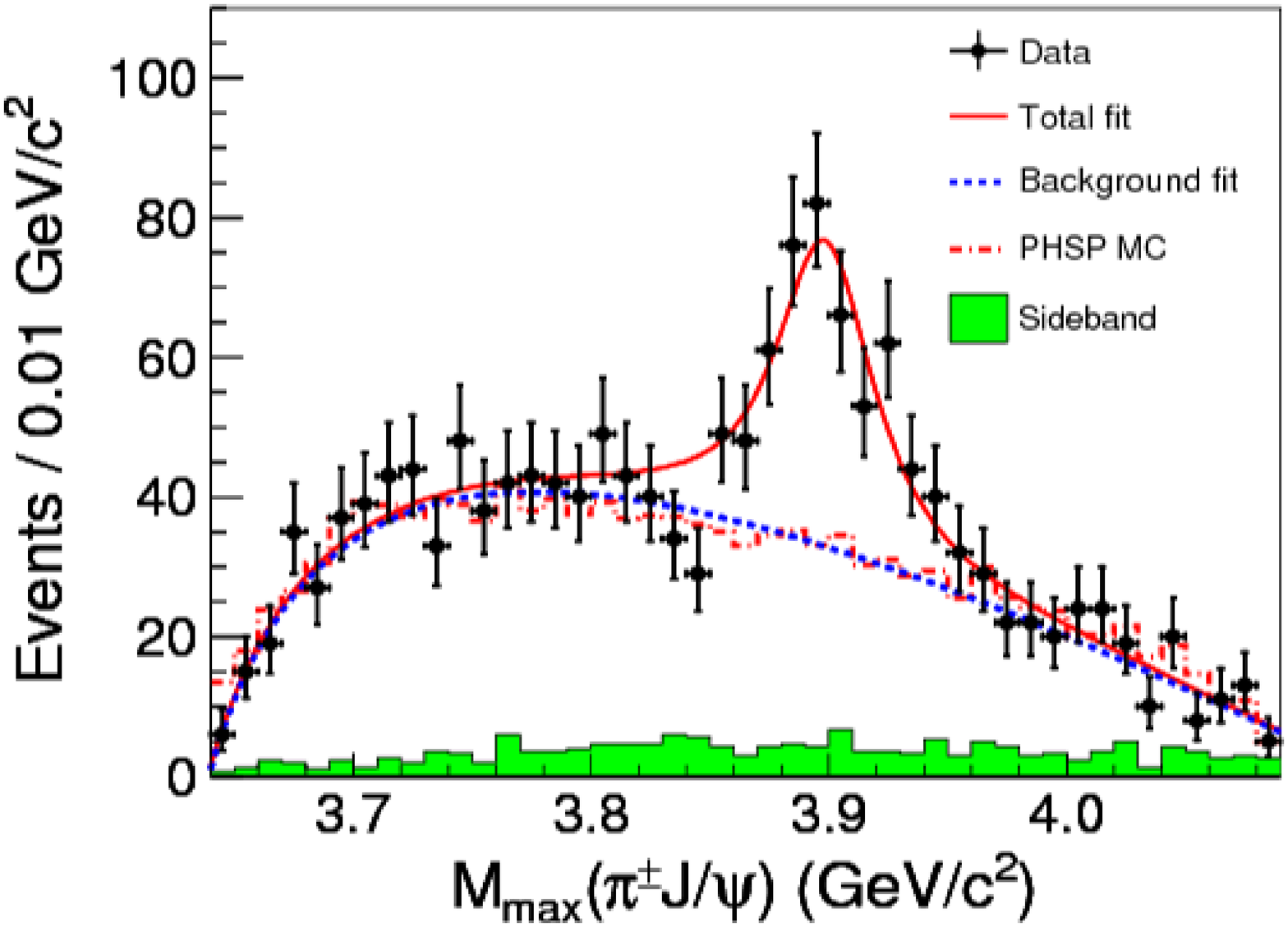}
\includegraphics[height=3.0in]{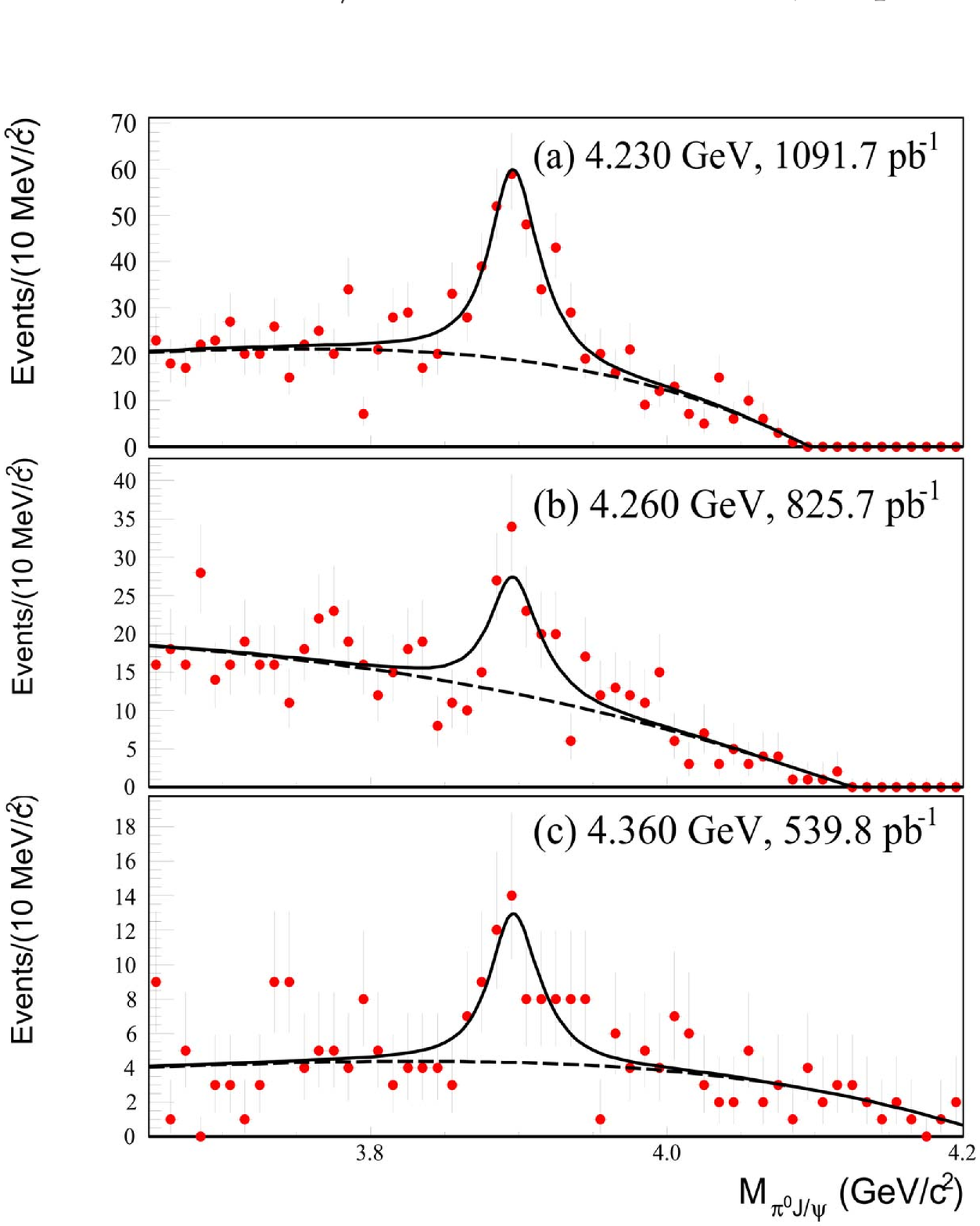}
\caption{$Z_c(3900)\to \pi J/\psi$ production in $e^+e^- \to \pi \pi J/\psi$ processes. 
The left plot shows charged channels~\cite{Ablikim:2013mio}, while the neutral mode~\cite{BESIII:2015kha}
is shown by the right plot. 
}
\label{fig:Zc3900pmz}
\end{figure}

Figure~\ref{fig:Zc3900pmz} shows the $Z_c(3900)^{\pm}$, identified in $\pi^{\pm} J/\psi$ decays (left),
found in the reaction $e^+ e^- \to \pi^+ \pi^- J/\psi$ at the center-of-mass energy, $E_{CM} = 4.26$ GeV, 
using a 525 pb$^{-1}$ data sample~\cite{Ablikim:2013mio}.
The plot on the right shows its isospin-partner,  $Z_c(3900)^0 \to \pi^{0} J/\psi$,
found in the reaction $e^+ e^- \to \pi^0 \pi^0 J/\psi$, at three energy points: 
$E_{CM}$ = 4.23, 4.26, and 4.36 GeV, with respective luminosities of 1097.1, 825.7 and 539.8 pb$^{-1}$~\cite{BESIII:2015kha}. 
The mass and width of the $\pi J/\psi$ structure in each case is determined from the fit, using a Breit-Wigner (BW) line shape 
combined with a non-resonant background contribution.  
Table~\ref{tab:Zc3900_mw} lists corresponding masses and widths. 

\subsection{$Z_c \to to D \overline{D}^{\star}$}

\begin{figure}[htb]
\centering
\includegraphics[height=2.0in]{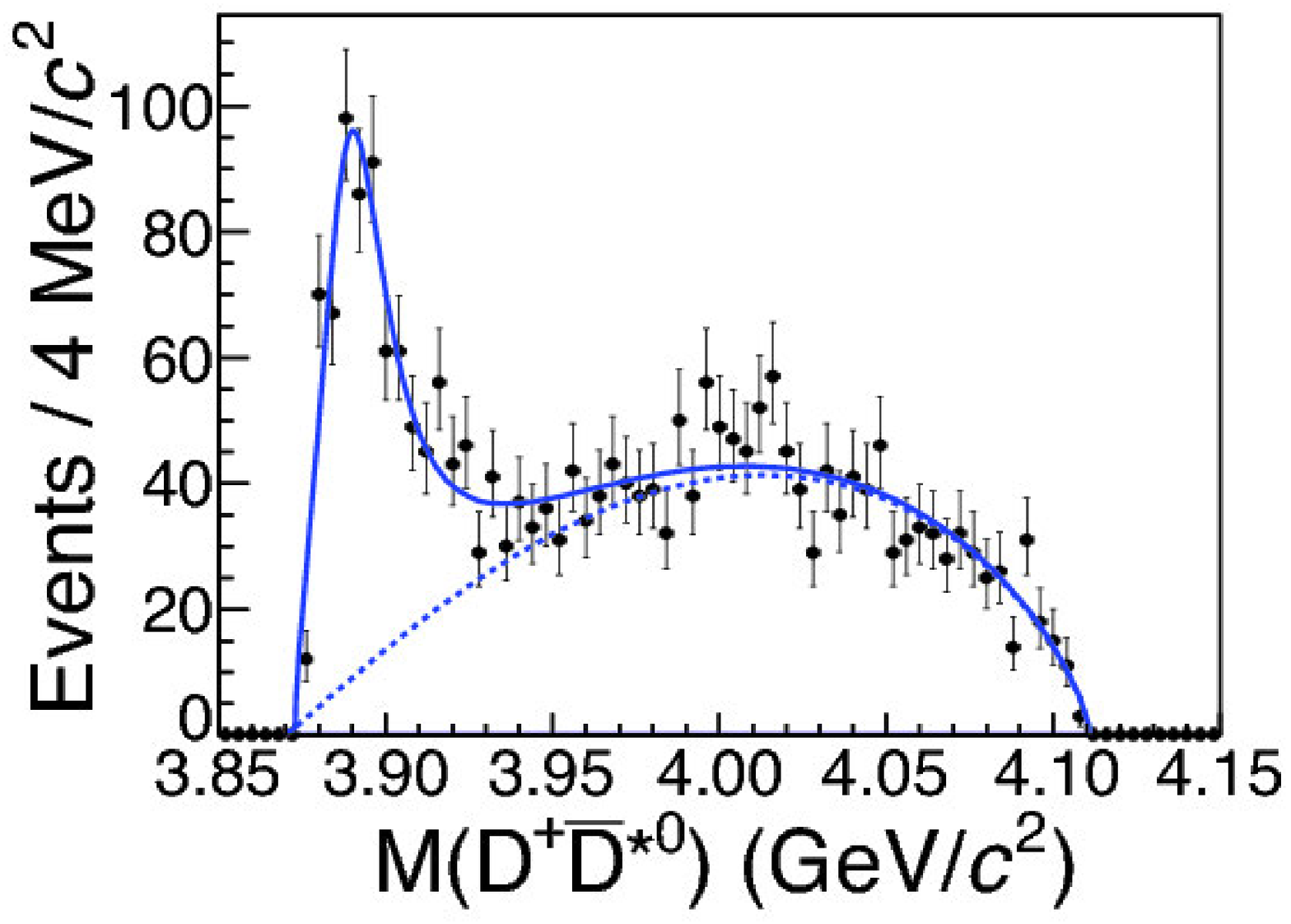}
\includegraphics[height=2.0in]{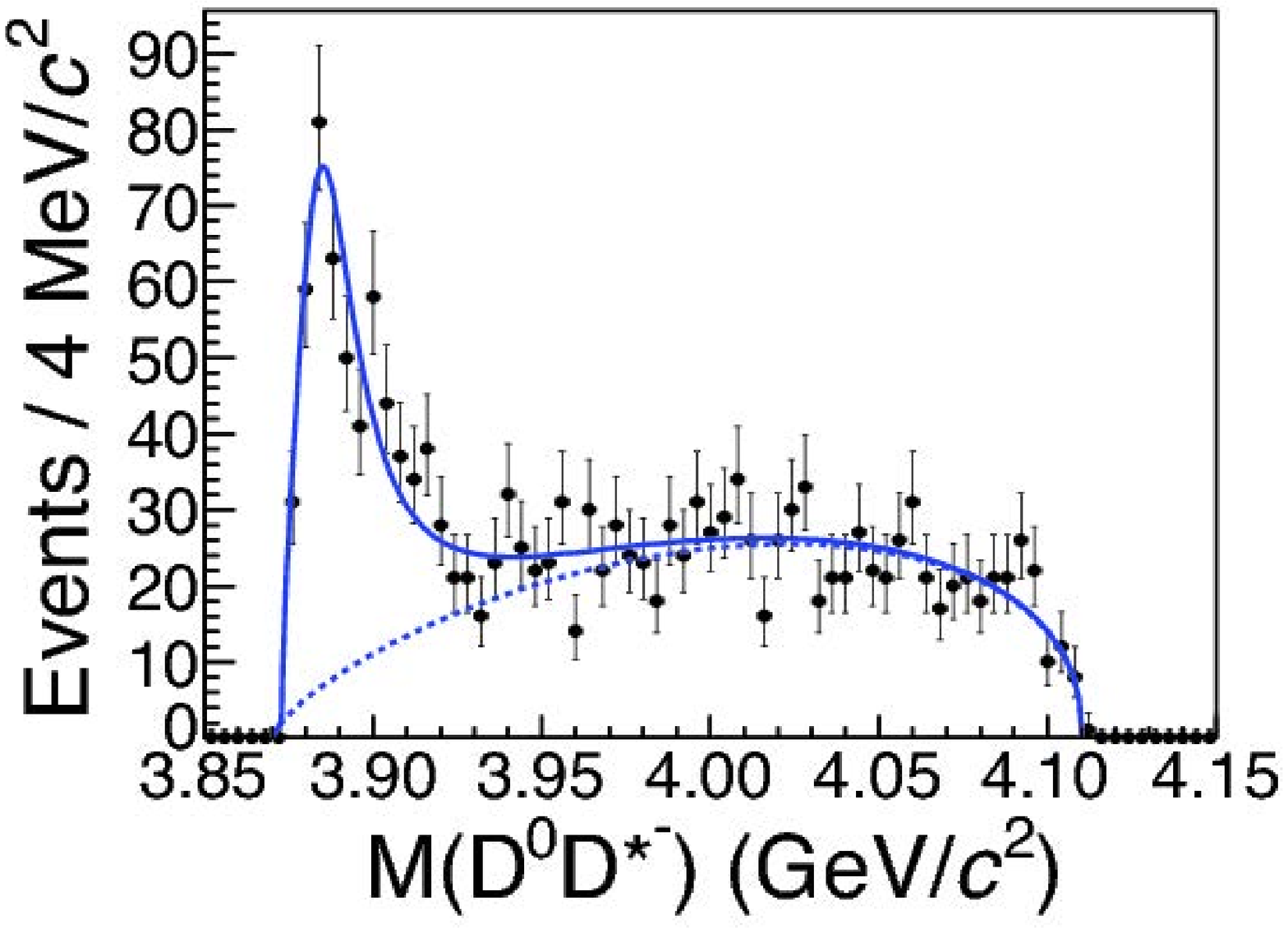}
\caption{ Structures identified as $Z_c(3885) \to D \overl{D}^{\star}$ when one 
charmed meson is found~\cite{Ablikim:2013xfr} are evident. 
}
\label{fig:Zc3885_1DT}
\end{figure}

The $Z_c(3900)$ is close to the threshold for producing two charmed mesons, $D$ and $\overl{D}^{\star}$.
Using a 525 pb$^{-1}$ data sample collected at $E_{CM}$ = 4.26 GeV,
BESIII studied the processes $e^+e^- \to \pi^{\pm} (D \overl{D}^{\star})^{\pm}$, in which one charged 
pion and one $D$ meson are identified, requiring that the missing mass is consistent 
with a $\overl{D}^*$~\cite{Ablikim:2013xfr}, in so-called single D-tag procedure (ST).  
Resulting mass distributions of $D \overl{D}^{\star}$ pairs are shown in Figure~\ref{fig:Zc3885_1DT}.
In both decay channels, structures that cannot be explained by underlying background are evident, 
which we labeled as $Z_c(3885)$.

\begin{figure}[htb]
\centering
\includegraphics[height=2.0in]{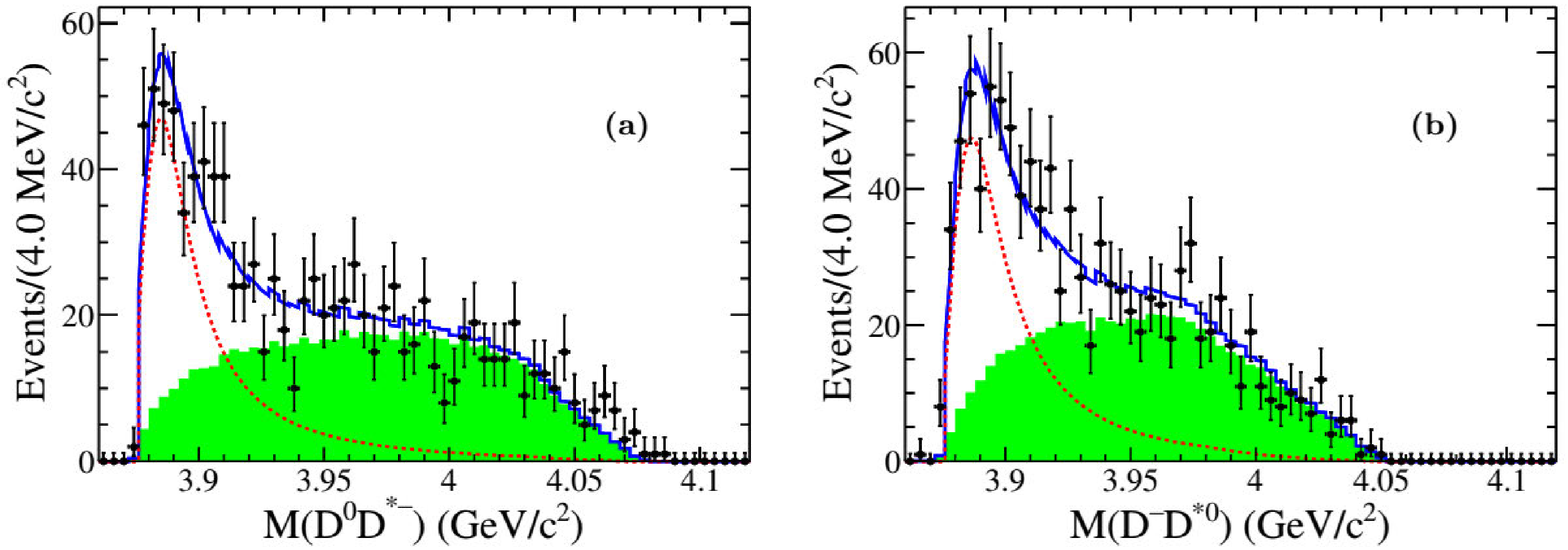}
\includegraphics[height=2.0in]{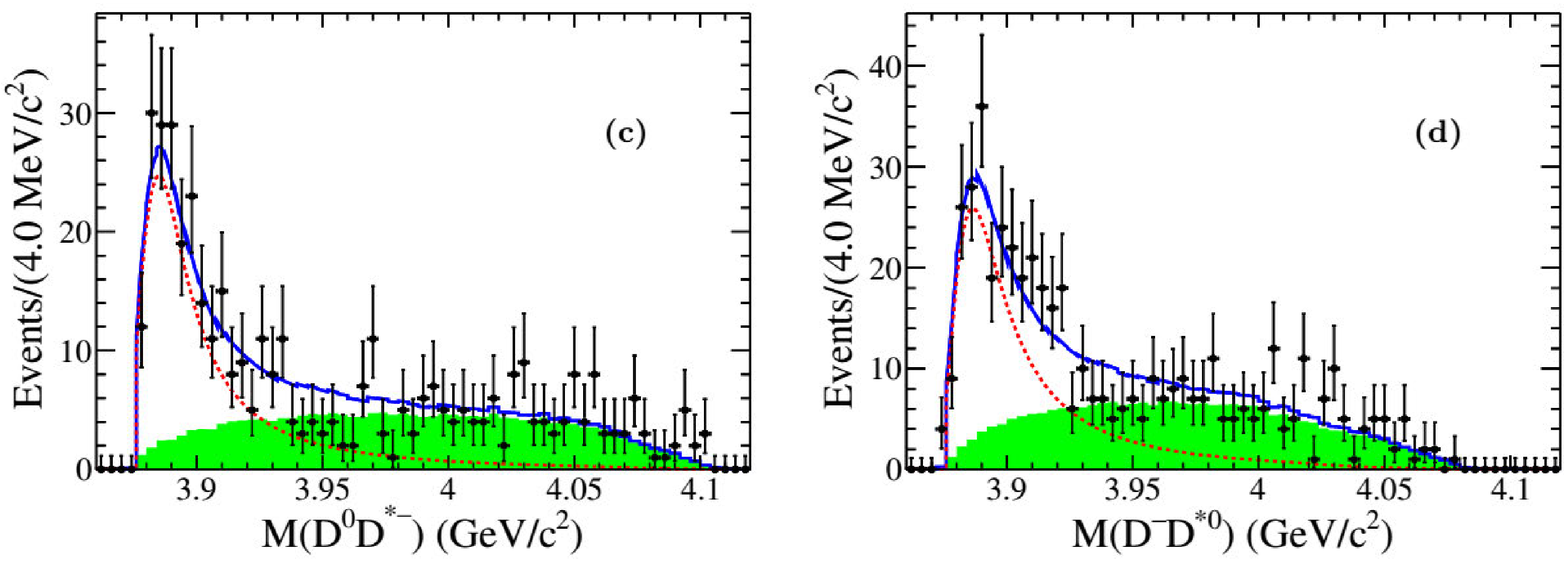}
\caption{ $Z_c(3883) \to D \overl{D}^{\star}$ structures that were evident when one $D$-meson is found, 
Fig.~\ref{fig:Zc3885_1DT}, are also present when  
both $D$ mesons are identified~\cite{Ablikim:2013xfr}.
Top plots are obtained at $E_{CM} = 4.23$~GeV, while bottom plots correspond to $E_{CM} = 4.26$~GeV.
}
\label{fig:Zc3885_2DT}
\end{figure}

Using data samples collected at $E_{CM}$ = 4.23 and 4.26 GeV, with respective luminosities of 1092 and 826 pb$^{-1}$, 
BESIII studied the same reaction with the requirement that both charmed mesons are fully reconstructed~\cite{Ablikim:2015swa},
in a double D-tag (DT) analysis.
Resulting mass distributions, shown in Fig.~\ref{fig:Zc3885_2DT},
exhibit the same features observed in the ST analysis. 
The pole positions, extracted from mass-dependent BW-line shapes are 
provided in Table~\ref{tab:Zc3900_mw}, where one can see that the DT-procedure reduced 
the errors.
The mass difference with respect to the $Z_c(3900)$ falls outside the error boundaries, 
indicating that the $Z_c(3885)$ and $Z_c(3900)$ might be different objects.   
Angular analysis, favor the spin-parity assignment 
of $J^P = 1^{+}$ for this structure, in both ST and DT procedures.

\subsection{$Z_c \to \omega \pi$}
\begin{figure}[htb]
\centering
\includegraphics[height=2.0in]{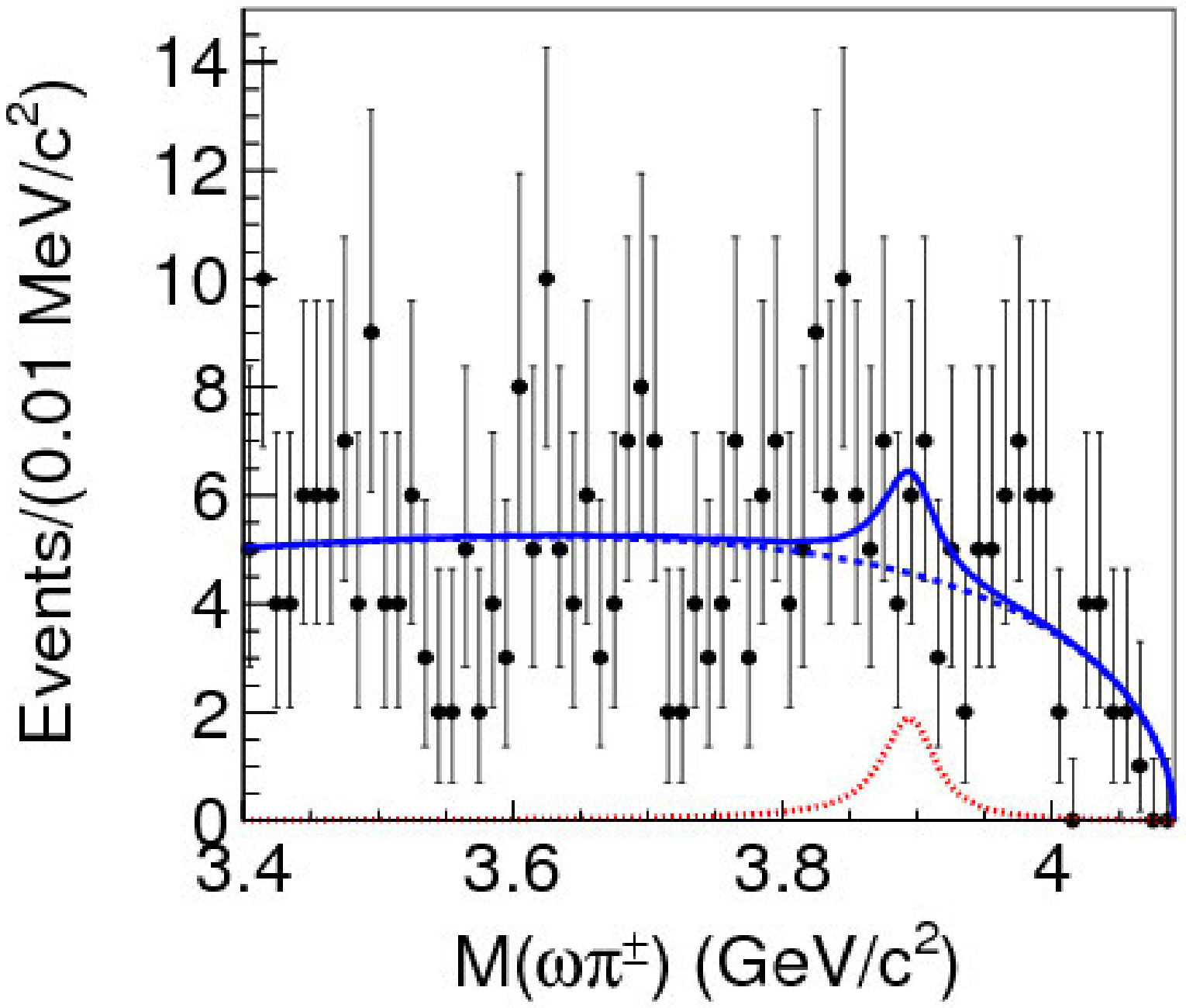}
\includegraphics[height=2.0in]{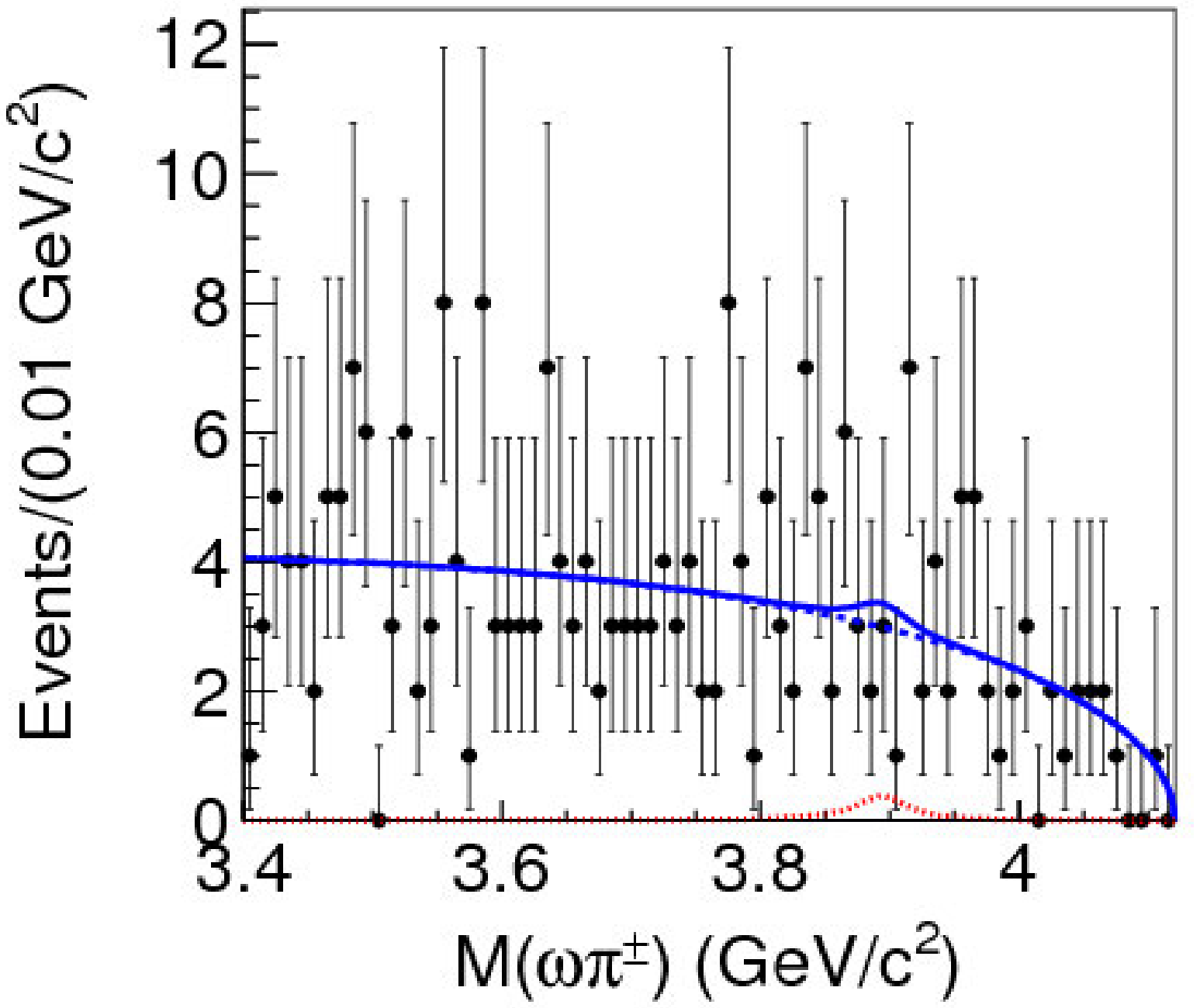}
\caption{ $\omega \pi$ mass spectra,  
obtained from the reaction $e^+e^- \to \omega \pi^+ \pi^-$ at two energy points: 
$E_{CM}$ = 4.23~GeV (left) and $E_{CM}$ = 4.26~GeV (right)~\cite{Ablikim:2015cag}.
}
\label{fig:Zc3900_opi}
\end{figure}

BESIII also searched for the production of $Z_c$ states in the reaction 
$e^+e^- \to \omega \pi^+ \pi^-$ at $E_{CM}$ = 4.23 and 4.26 GeV~\cite{Ablikim:2015cag}. 
Fig.~\ref{fig:Zc3900_opi} shows the $\omega \pi^{\pm}$ mass distribution, at $E_{CM}$ = 4.23~GeV (left) 
and at $E_{CM}$ = 4.26~GeV (right). There is no significant $Z_c(3900) \to \omega \pi$ signal, 
and the upper limits 
oh the Born cross section are determined to be 0.26 and 0.18 pb, at $E_{CM}$ = 4.23 and 4.26 GeV, respectively.  

\subsection{ $Z_c \to \pi h_c$}

\begin{figure}[htb]
\centering
\includegraphics[height=2.0in]{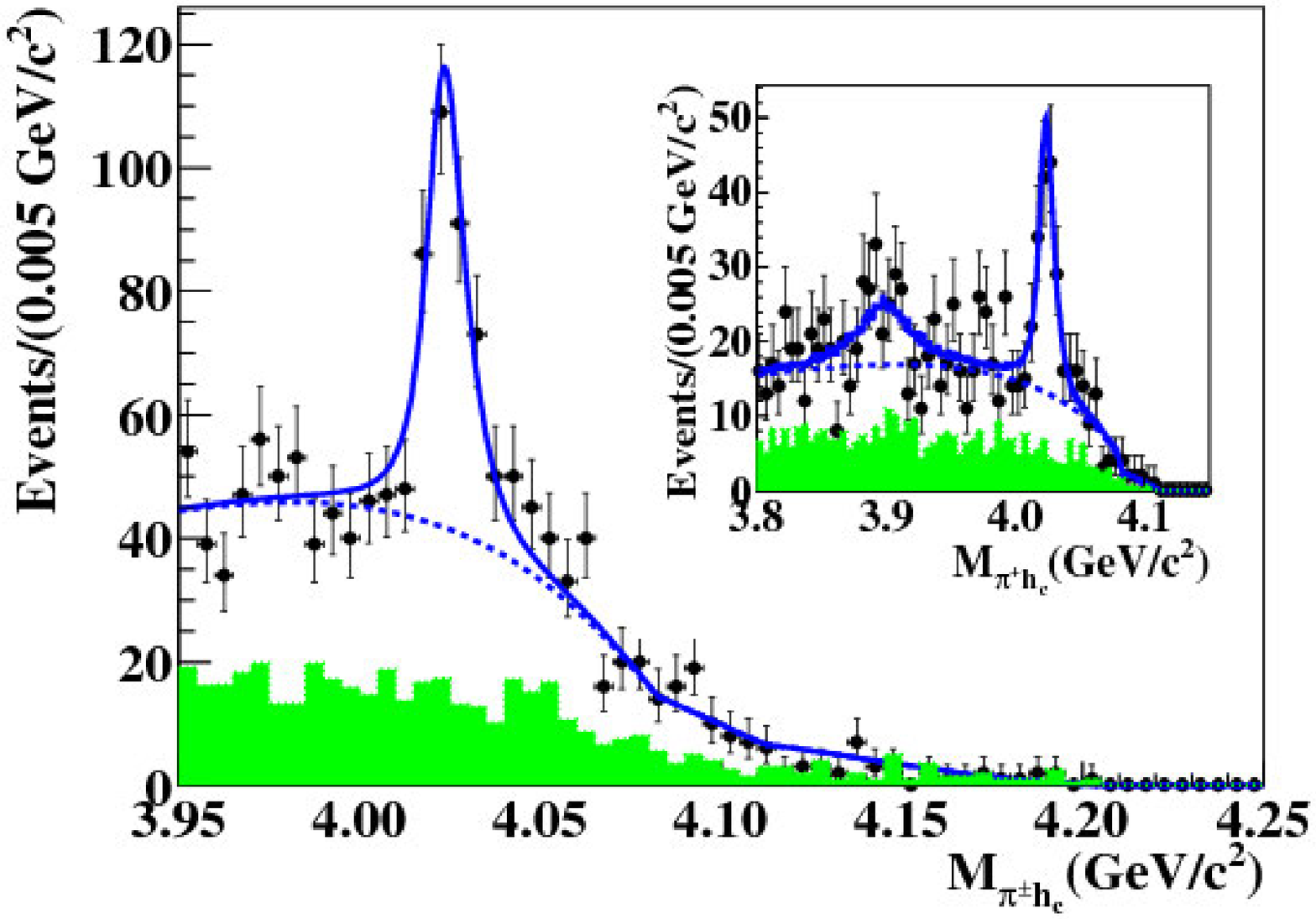}
\includegraphics[height=2.0in]{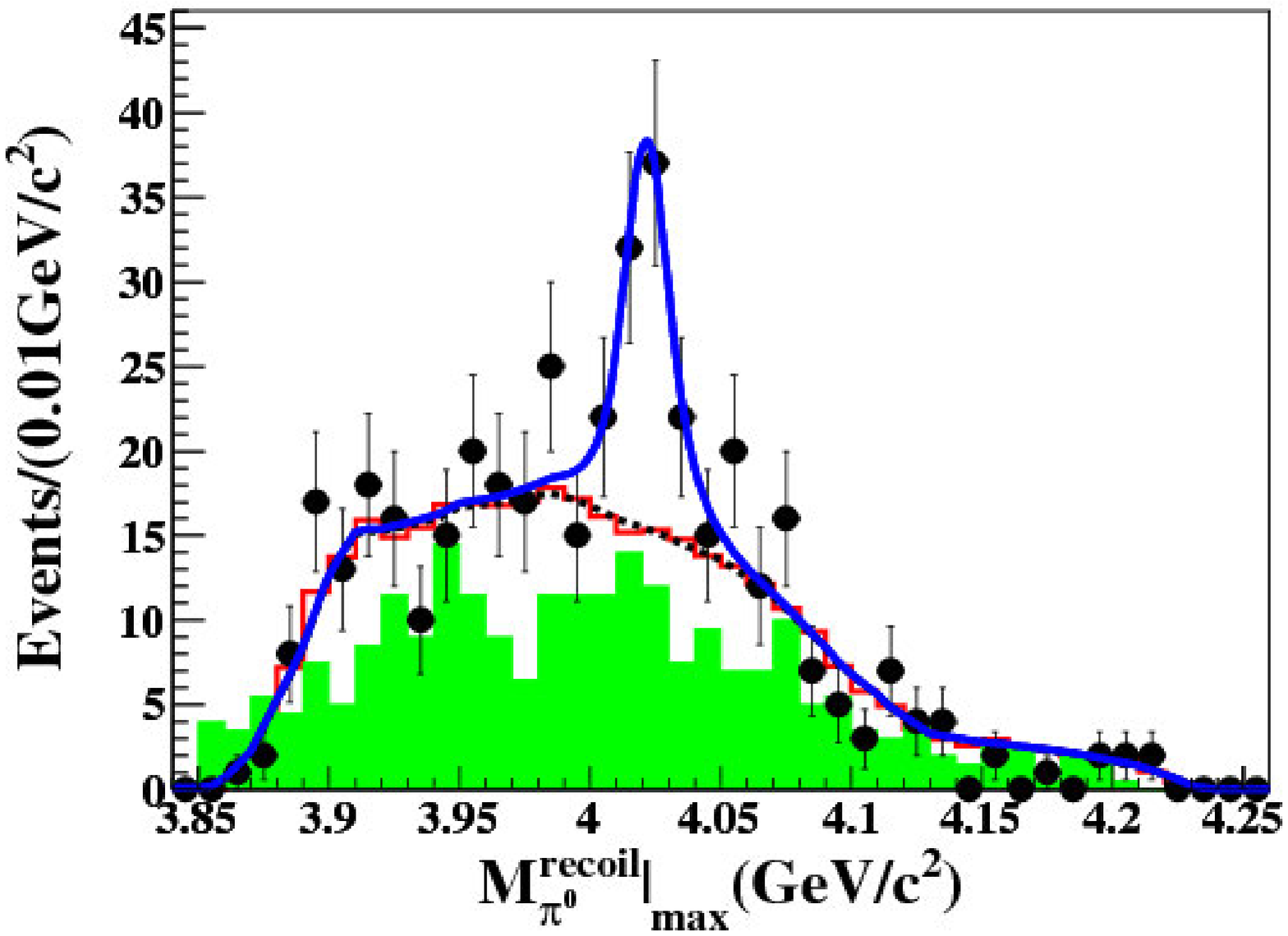}
\caption{Production of the $Z_c(4020)\to \pi h_c$ in $e^+e^- \to \pi \pi h_c$ processes, 
at $E_{CM}$ = 4.23, 4.26, and 4.36~GeV, evident in charged~\cite{Ablikim:2013wzq} (left) 
and neutral ~\cite{Ablikim:2014dxl} (right) modes. 
The inset shows a fit that assumes the $Z_c(3900) \to \pi h_c$ production, 
at $E_{CM}$ = 4.23 and 4.26~GeV.
}
\label{fig:Zc4020_pmz}
\end{figure}

BESIII analyzed the  reaction 
$e^+e^- \to \pi^+ \pi^- h_c$, at $E_{CM}$ = 4.23, 4.26 and 4.36~GeV~\cite{Ablikim:2013wzq}.
Corresponding $\pi^{\pm} h_c$ invariant mass distribution, when all energy points are combined, is shown by the left plot of Fig.~\ref{fig:Zc4020_pmz}.
The $Z_c(4020)^{\pm}$ signal has 8.9$\sigma$ significance, and 
the inset represents a search for the $Z_c(3900) \to \pi h_c$, at $E_{CM}$ = 4.23 and 4.26~GeV. 

BESIII investigated the reaction $e^+e^- \to \pi^0 \pi^0 h_c$, 
at $E_{CM}$ = 4.23, 4.26 and 4.36~GeV~\cite{Ablikim:2014dxl}.  
The right plot in Fig.~\ref{fig:Zc4020_pmz} shows the evidence for the neutral partner of the $Z_c(4020)^{\pm}$, 
in the mass distribution of a recoiling $\pi^0$. 
Each structure, observed in the charged and neutral mode, is assumed to have a BW-line shape, 
and resulting masses and widths are given in Table~\ref{tab:Zc3900_mw}. 
In the case of the $Z_c(4020)^0$ fit, the width is fixed to the value obtained analyzing 
charged decays.  

\subsection{$Z_c \to to D^{\star} \overline{D}^{\star}$}

\begin{figure}[htb]
\centering
\includegraphics[height=2.0in]{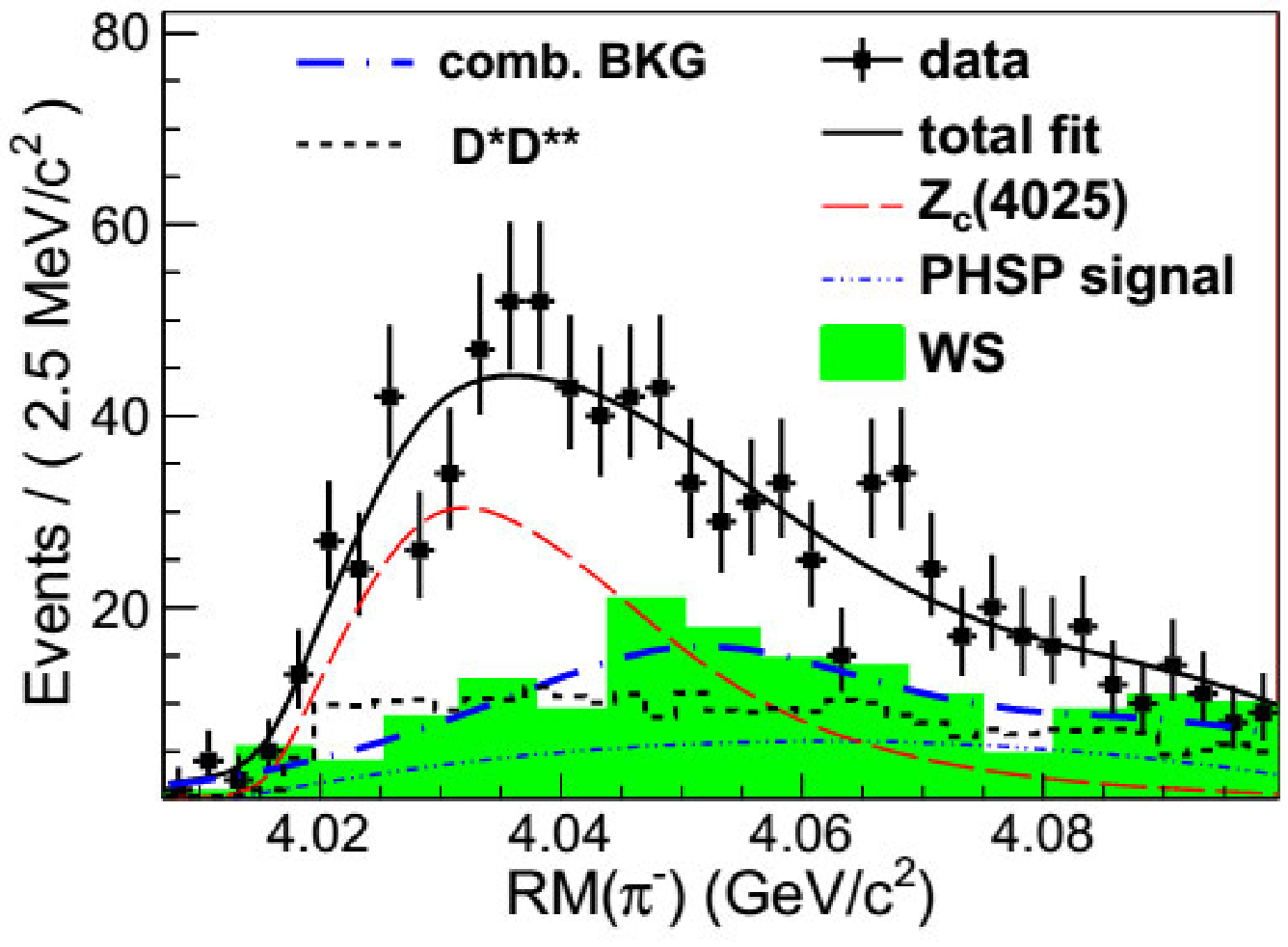}
\includegraphics[height=2.0in]{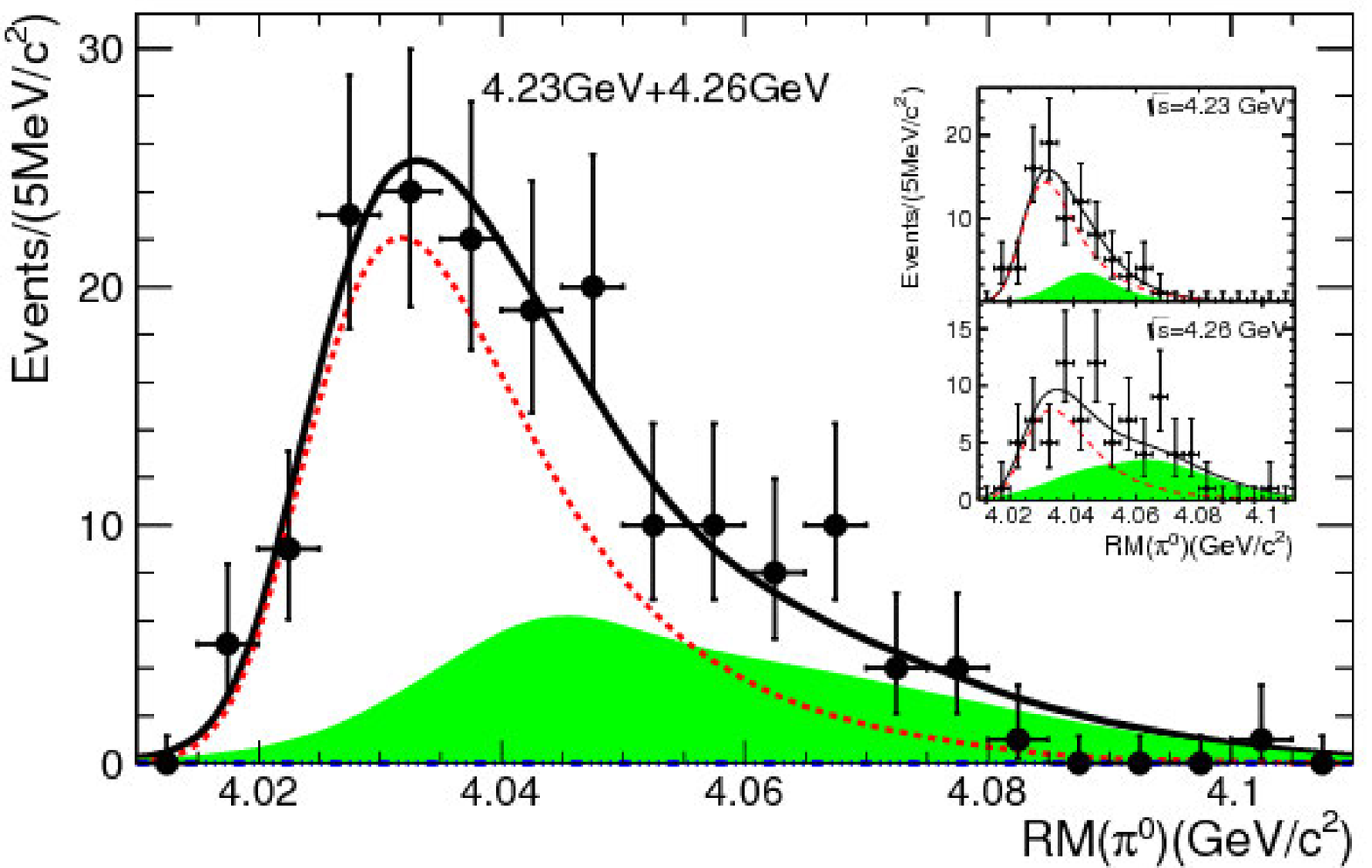}
\caption{Production of the $Z_c(4025)\to (D^* \overl{D}^*)^{\pm}$ in 
the process $e^+e^- \to \pi^{\pm} (D^{\star} \overl{D}^{\star})^{\mp}$, 
at  $E_{CM}$ = 4.26~GeV~\cite{Ablikim:2013emm} (left). 
The $Z_c(4025)\to (D^{\star} \overl{D}^{\star})^{0}$ signal can be observed in the 
right plot, obtained in the process $e^+e^- \to \pi^{0} (D^{\star} \overl{D}^{\star})^{0}$,
at $E_{CM}$ = 4.23 and 4.26~GeV~\cite{Ablikim:2015vvn}.
}
\label{fig:Zc4025_DsDs}
\end{figure}

Similarly to the case of the $Z_c(3900)$, the mass of the $Z_c(4020)$ is near
the threshold for $D^{\star} \overl{D}^{\star} $ production.
BESIII studied two reactions: $e^+e^- \to (D^{\star} \overl{D} ^{\star})^{\pm} \pi^{\mp}$,
at $E_{CM}$ = 4.26~GeV~\cite{Ablikim:2013emm}, 
and $e^+e^- \to (D^{\star} \overl{D} ^{\star})^{0} \pi^{0}$,
at $E_{CM}$ = 4.23 and 4.26~GeV~\cite{Ablikim:2015vvn}. 
The respective luminosities at two energy points are 1092 and  826 pb$^{-1}$.
Fig.~\ref{fig:Zc4025_DsDs} shows the $\pi^-$ (left) and $\pi^0$ (right) 
recoil mass distributions. A structure that couples to $D^{\star} \overl{D^{\star}}$ 
is evident, in both charged and neutral decays. The mass and width are extracted in each case using a BW-line shape, 
and corresponding values are listed in Table~\ref{tab:Zc3900_mw}. The difference in 
values with respect to the $Z_c(4020)$, which couples to $\pi h_c$, warrants labeling this 
structure as a separate state: $Z_c(4025)$.

\begin{table}[t]
\begin{center}
\begin{tabular}{l|cc|c}  

Decay &  Mass [MeV/$c^2$] &  Width [MeV]  & Ref.  \\ 
\hline
$Z_c(3900)^{\pm} \to \pi^{\pm} J/\psi$ & 
 3899.0$\pm$3.6$\pm$4.9 & 46$\pm$10$\pm$20 &  \cite{Ablikim:2013mio} \\
$Z_c(3900)^{0} \to \pi^{0} J/\psi$ & 
3894.8$\pm$2.3$\pm$3.2 & 29.6$\pm$8.2$\pm$8.2 & \cite{BESIII:2015kha} \\ 

\hline
$Z_c(3885)^{\pm} \to (D \overl{D}^*)^{\pm}$ (ST) & 
  3883.9$\pm$1.5$\pm$4.2 & 24.8$\pm$3.3$\pm$11.0 &  ~\cite{Ablikim:2013xfr} \\
$Z_c(3885)^{\pm} \to (D \overl{D}^*)^{\pm}$ (DT) & 
  3881.7$\pm$1.6$\pm$2.1 & 26.6$\pm$2.0$\pm$2.3 & ~\cite{Ablikim:2015swa} \\

\hline
$Z_c(4020)^{\pm} \to \pi^{\pm} h_c$ &  
4022.9$\pm$0.8$\pm$2.7 & 7.9$\pm$2.7$\pm$2.6 & \cite{Ablikim:2013wzq} \\
$Z_c(4020)^{0} \to \pi^{0} h_c$ &  
  4023.9$\pm$2.2$\pm$3.8 & fixed &  \cite{Ablikim:2014dxl} \\
\hline

$Z_c(4025)^{\pm} \to (D^* \overl{D} ^*)^{\pm}$ & 
4026.3$\pm$2.6$\pm$3.7 & 24.8$\pm$5.6$\pm$7.7 &  \cite{Ablikim:2013emm} \\
$Z_c(4025)^{0} \to (D^* \overl{D} ^*)^{0}$ & 
4025.5$^{+2.0}_{-4.7}\pm3.1$ & 23.0$\pm$6.0$\pm$1.0 & \cite{Ablikim:2015vvn} \\

\hline
\end{tabular}
\caption{Masses and widths of exotic meson candidates studied by BESIII.}
\label{tab:Zc3900_mw}
\end{center}
\end{table}



\section{Relation between $XYZ$ states}

\begin{figure}[htb]
\centering
\includegraphics[height=2in]{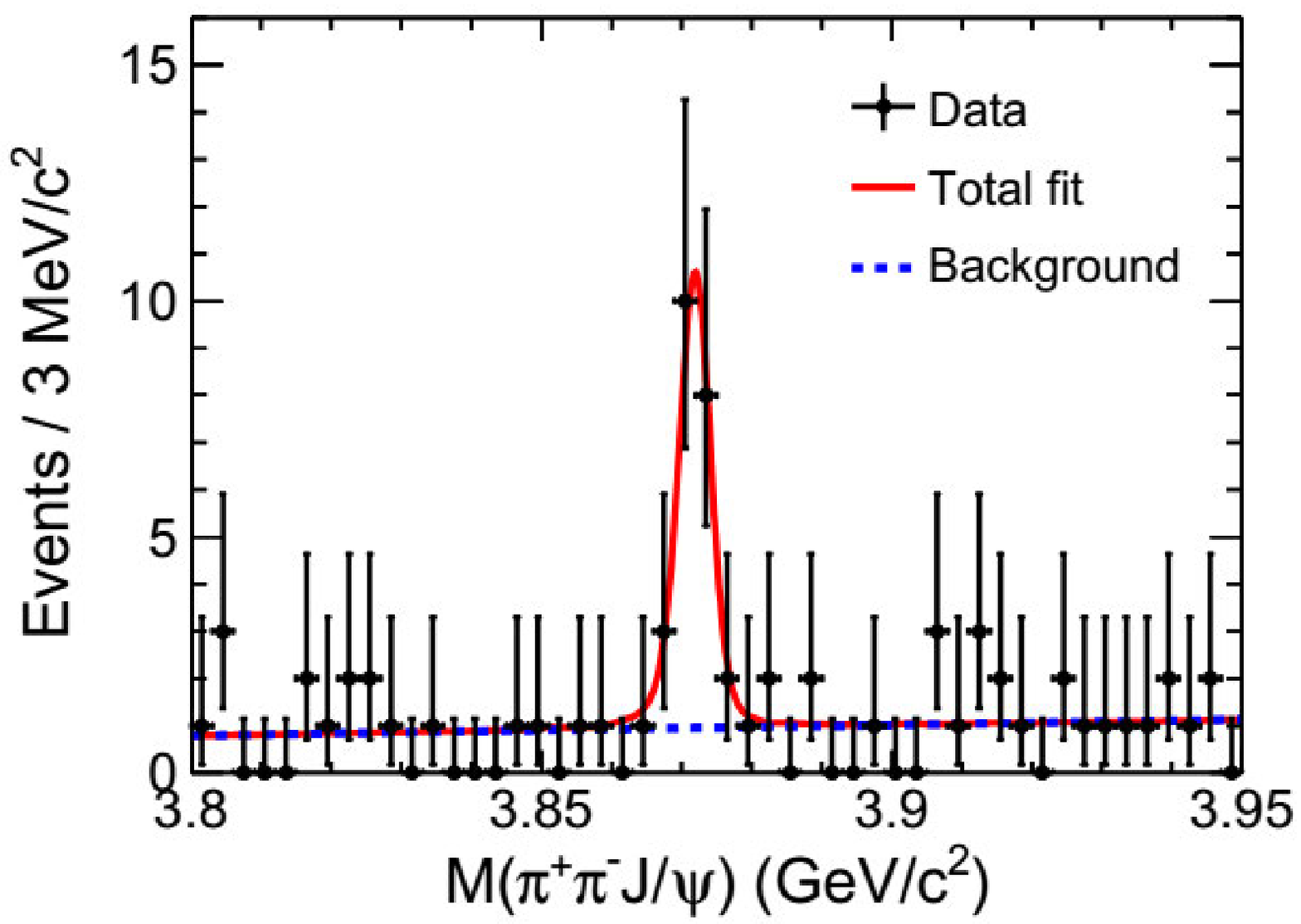}
\includegraphics[height=2.0in]{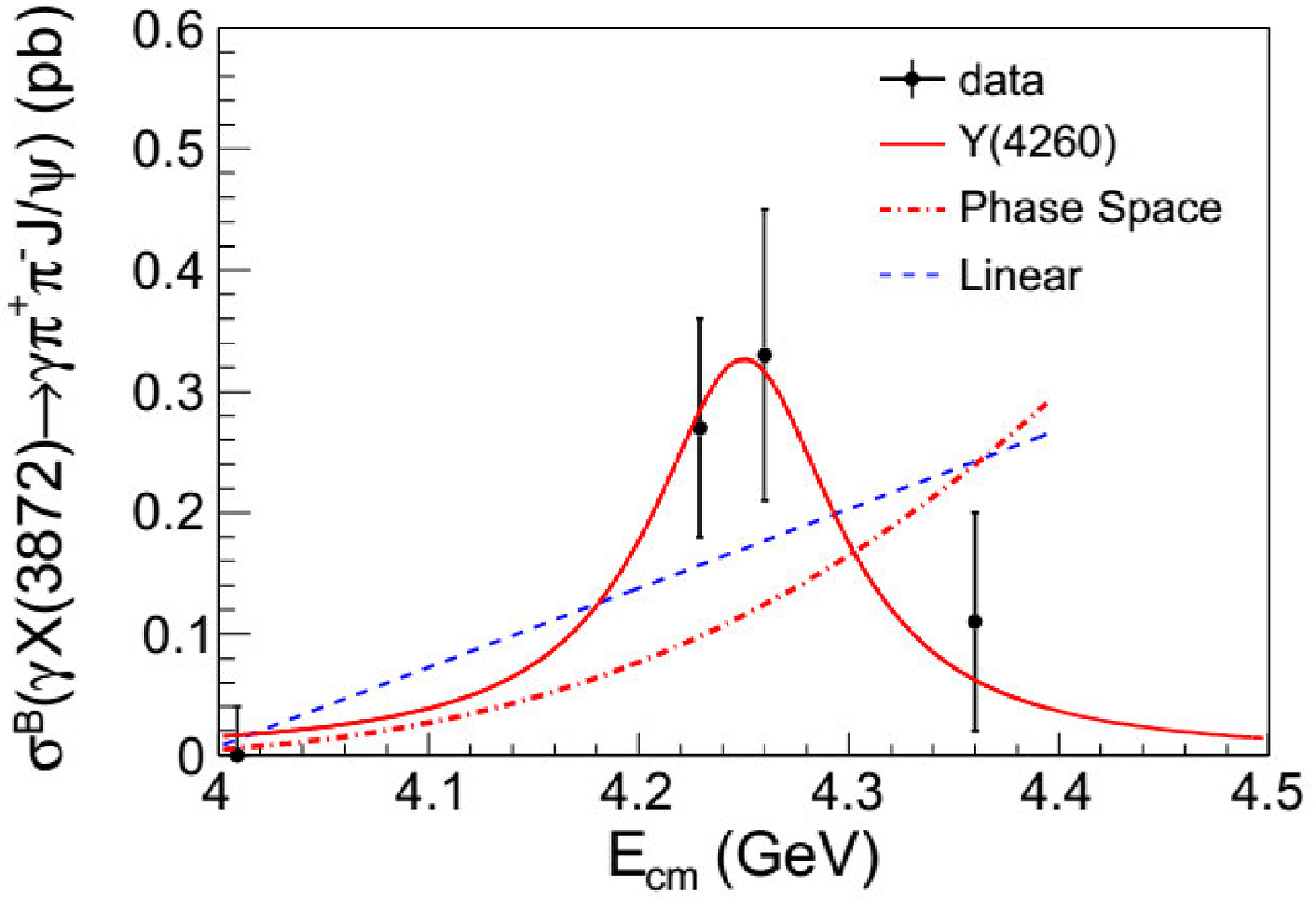}
\caption{X(3872) identified in the $\pi^+ \pi^- J/\psi$ decays (left). 
The cross section for the reaction 
$e^+ e^- \to \gamma \pi^+ \pi^- J/\psi$ as a function of energy (right)~\cite{Ablikim:2013dyn}.
}
\label{fig:Y_gX}
\end{figure}

Two notorious meson candidates that do not fit into the quark-anti-quark model are 
$X(3872)$ and $Y(4260)$, discovered by Belle~\cite{Choi:2003ue} and BaBar~\cite{Aubert:2005rm},
respectively.
BESIII is also able to investigate their properties, by looking into 
the reaction $e^+ e^- \to \gamma \pi^+ \pi^- J/\psi$ in the energy region 
between 4.2 and 4.6 GeV. 
Fig.~\ref{fig:Y_gX} shows the invariant mass of $\pi^+ \pi^- J/\psi$ candidates (left) 
where a clear $X(3872)$ signal can be seen. The right plot shows the cross section as 
a function of energy, compared with the line shape of the $Y(4260)$  production~\cite{Ablikim:2013dyn}.
This plot indicates that there is a connection between the two states: 
$Y(4260) \to \gamma X(3872)$, which has not been reported before. 
BESIII is also looking into possible connections between the $Y(4260)$, and other Y states,
to $Z_c$ particles.

\clearpage
\section{Summary}

BESIII is making progress in shedding light on properties of newly discovered exotic meson 
candidates. Two states that BESIII observed decaying to two different pion-charmonium 
combinations, namely the $Z_c(3900)$ and $Z_c(4020)$, are both established as
isospin triplets. 
There are also two structures that couple to  
$D D^{\star}$  and $D^{\star} D^{\star}$ pairs, namely  the $Z_c(3885)$ and $Z_c(4025)$.
It is possible that these are the same objects as the $Z_c(3900)$ and $Z_c(4020)$, respectively, 
because of their proximity to corresponding $DD$-thresholds. In addition, the $Z_c(4025)$ is also 
established as an isospin triplet, and BESIII is looking for the neutral partner of the $Z_c(3885)$.
The spin analysis is also needed in order to establish, for example, that the $Z_c(3900)$ is the same as the $Z_c(3885)$.
The evidence for the $Z_c(3900) \to \pi h_c$ and $Z_c(3900) \to \omega \pi$ 
decays is weak. However, BESIII observed the production of the $X(3872)$, 
in a process that is consistent with the radiative transition: $Y(4260) \to \gamma X(3872)$,
something that has not been observed before. 
Hopefully, all this information should help 
theorists working on providing explanations for these intriguing meson structures.


\Acknowledgements
We would like to thank all participants of CHARM 2015 in making this meeting a success. We also hope that 
all Proceedings contributions will be submitted on-time.



\begin{thebibliography}{99}


\bibitem{Fritzsch:1973pi} 
  H.~Fritzsch, M.~Jell-Mann and H.~Leutwyler,
  Phys.\ Lett.\ B {\bf 47}, 365 (1973).

\bibitem{Barnes:2005pb} 
  T.~Barnes, S.~Godfrey and E.~S.~Swanson,
  Phys.\ Rev.\ D {\bf 72}, 054026 (2005)

\bibitem{Brambilla:2014jmp} 
  N.~Brambilla {\it et al.},
  Eur.\ Phys.\ J.\ C {\bf 74}, no. 10, 2981 (2014)


\bibitem{Brambilla:2010cs} 
  N.~Brambilla {\it et al.},
  Eur.\ Phys.\ J.\ C {\bf 71}, 1534 (2011)

\bibitem{Ablikim:2013mio} 
  M.~Ablikim {\it et al.} [BESIII Collaboration],
  Phys.\ Rev.\ Lett.\  {\bf 110}, 252001 (2013)

\bibitem{Ablikim:2013wzq} 
  M.~Ablikim {\it et al.} [BESIII Collaboration],
  Phys.\ Rev.\ Lett.\  {\bf 111}, no. 24, 242001 (2013)




\bibitem{BESIII:2015kha} 
  [BESIII Collaboration],
  arXiv:1506.06018 [hep-ex].



\bibitem{Ablikim:2013xfr} 
  M.~Ablikim {\it et al.} [BESIII Collaboration],
  Phys.\ Rev.\ Lett.\  {\bf 112}, no. 2, 022001 (2014)

\bibitem{Ablikim:2015swa} 
  M.~Ablikim {\it et al.} [BESIII Collaboration],
  arXiv:1509.01398 [hep-ex].

\bibitem{Ablikim:2015cag} 
  M.~Ablikim {\it et al.} [BESIII Collaboration],
  Phys.\ Rev.\ D {\bf 92}, no. 3, 032009 (2015)




\bibitem{Ablikim:2014dxl} 
  M.~Ablikim {\it et al.} [BESIII Collaboration],
  Phys.\ Rev.\ Lett.\  {\bf 113}, no. 21, 212002 (2014)


\bibitem{Ablikim:2013emm} 
  M.~Ablikim {\it et al.} [BESIII Collaboration],
  Phys.\ Rev.\ Lett.\  {\bf 112}, no. 13, 132001 (2014)
\bibitem{Ablikim:2015vvn} 
  M.~Ablikim {\it et al.} [BESIII Collaboration],
  arXiv:1507.02404 [hep-ex].


%
\bibitem{Choi:2003ue} 
  S.~K.~Choi {\it et al.} [Belle Collaboration],
  Phys.\ Rev.\ Lett.\  {\bf 91}, 262001 (2003)

\bibitem{Aubert:2005rm} 
  B.~Aubert {\it et al.} [BaBar Collaboration],
  Phys.\ Rev.\ Lett.\  {\bf 95}, 142001 (2005)


\bibitem{Ablikim:2013dyn} 
  M.~Ablikim {\it et al.} [BESIII Collaboration],
  Phys.\ Rev.\ Lett.\  {\bf 112}, no. 9, 092001 (2014)


\end{thebibliography}
\end{document}